\documentclass[cits]{PoS}
\pdfoutput=1

\usepackage{amsmath}

\DeclareMathOperator{\sgn}{sgn}
\DeclareMathOperator{\re}{Re}

\DeclareMathOperator{\myspan}{span}

\DeclareMathOperator{\diag}{diag}

\newcommand{\Dov}{D_\text{ov}}
\newcommand{\Dw}{D_\text{w}}
\newcommand{\kout}{k}
\newcommand{\kin}{\ell}
\newlength{\figwidth}

\title{A nested Krylov subspace method for the overlap operator}

\ShortTitle{A nested Krylov subspace method for the overlap operator}

\dedicated{Supported by the DFG collaborative research center SFB/TR-55 "Hadron Physics from Lattice QCD".}

\author{\speaker{Jacques C.R. Bloch}
        \\
       University of Regensburg\\
       E-mail: \email{jacques.bloch@physik.uni-regensburg.de}}
\author{Simon Heybrock
        \\
       University of Regensburg\\
       E-mail: \email{simon.heybrock@physik.uni-regensburg.de}}

\abstract{
 We present a novel method to compute the overlap 
Dirac operator at zero and nonzero quark chemical potential. To approximate 
the sign function of large, sparse matrices, standard methods project 
the operator on a much smaller Krylov subspace, on which the matrix 
function is computed exactly. However, for large lattices this 
subspace can still be too large for an efficient calculation of the 
sign function. 
The idea of the new method is to nest Krylov subspace approximations 
by making a further projection on an even smaller 
subspace, which is then small enough to compute the sign function 
efficiently, and this without any noticeable loss of numerical 
accuracy. 
We demonstrate the efficiency of the method both on 
Hermitian and non-Hermitian matrices.
}

\FullConference{The XXVII International Symposium on Lattice Field Theory\\
		 July 26-31, 2009\\
		 Peking University, Beijing, China}

\begin{document}

\section{The overlap operator and the sign function}

In this contribution we present an improved method to compute the sign function of a large sparse matrix, as necessitated by the overlap Dirac operator in lattice QCD. More details and results can be found in Ref.~\cite{S&J:2009}. The new method applies to Hermitian and non-Hermitian matrices and is therefore usable in QCD simulations at both zero and  nonzero quark chemical potential.

The overlap operator introduced by Neuberger and Narayanan \cite{Narayanan:1994gw,Neuberger:1997fp} and extended to nonzero quark chemical potential $\mu$ by Bloch and Wettig \cite{Bloch:2006cd} is defined as 
\begin{align}
        \Dov(\mu) &= 1+\gamma_5 \sgn\left(\gamma_5 \Dw(\mu)\right) ,
        \label{Dov}
\end{align}
with Wilson-Dirac operator $\Dw(\mu)$ and Wilson mass satisfying $-2 < m_\text{w} < 0$. The quark chemical potential is introduced in the Wilson-Dirac operator as prescribed by Hasenfratz and Karsch \cite{Hasenfratz:1983ba}.
The matrix $\gamma_5 \Dw $ is Hermitian when $\mu = 0$, but becomes non-Hermitian when $\mu \ne 0$, such that the overlap operator requires the computation of the sign of a general complex matrix.

For a generic function $f$ and a diagonalizable matrix $A = U \diag(\lambda_1, \cdots, \lambda_n) U^{-1} \in \mathbb{C}^{n\times n}$, with eigenvalues $\lambda_i$ and eigenvector matrix $U=[u_1|\dotsc|u_n]$, the matrix function $f(A)$ is given by the spectral definition
\begin{align}
    f(A) &= U \diag\left(f(\lambda_1), \ldots, f(\lambda_n) \right)
            U^{-1} .
\label{specdef}
\end{align}
In the non-Hermitian case the eigenvalues are generically complex and the sign function is defined as $\sgn(z) = \sgn(\re(z))$ \cite{Bloch:2007xi}.

\section{Approximate Krylov subspace solution}

For physically relevant lattice sizes the spectral definition \eqref{specdef} cannot be applied directly as the diagonalization of $A$ becomes too expensive in terms of CPU time and memory. 
A solution is to compute $f(A) x$, for $x\in\mathbb{C}^n$,  using a Krylov subspace approximation, rather than $f(A)$, as the results of such operations are typically needed by iterative solvers for linear systems and eigensystems.

To approximate $f(A) x$ in the Krylov subspace 
$\mathcal{K}_k(A,x) = \myspan(x,Ax,\dotsc,A^{k-1}x)$
 we first construct a basis for the subspace.
 In the Hermitian case the Lanczos method generates an orthonormal basis $V_k$ of $\mathcal{K}_k(A,x)$ using short recurrences.
In the non-Hermitian case an orthonormal basis can be constructed using the Arnoldi algorithm. 
However, as the method uses long recurrences, it becomes too expensive for large Krylov subspaces.
Instead, we will use the two-sided or biorthogonal Lanczos method (2sL), where the orthogonality of the basis is given up in order to recover short recurrences. 
Two short recurrence relations are used to construct mutually orthonormal bases $V_k$ and $W_k$, i.e., $W_k^\dagger V_k=\mathbf{1}_k$, for the Krylov subspaces of $A$ and $A^\dagger$, respectively. 

With these bases the Ritz approximation to 
$f(A)x$ is computed using 
\begin{align}
  f(A) x  \approx V_k f(H_k) V_k^\dagger x =  \lVert x\rVert V_k f(H_k) e_1^{(k)} ,
  \label{approximation}
\end{align}
where we chose $v_1=x/\lVert x\rVert$, $e_1^{(k)}$ is the first unit vector of $\mathbb{C}^k$, and
the $k\times k$ Ritz matrix $H_k$ is defined as $V_k^\dagger A V_k$ and $ W_k^\dagger A V_k$, for the Hermitian and non-Hermitian case, respectively.
The virtue of the Krylov subspace approximation is that a good accuracy can be reached for $k \ll n$, such that the computation of $f(A)$ is replaced by that of $f(H_k)$, which is of much smaller size. 
For the sign function, $\sgn(H_k)$ is typically computed using the Roberts-Higham matrix iteration \cite{Rob80}:
\begin{align}
 S_{n+1} &= \frac{1}{2}(S_n + S_n^{-1}) \qquad \text{with} \quad  S_0 = A ,
 \label{Roberts}
\end{align}
which converges quadratically to $\sgn(A)$.
The combination of Eq.\eqref{approximation} with Eq.~\eqref{Roberts} to compute $\sgn(H_k)$ will be called the non-nested method, in contrast to the nested method introduced in Sec.~\ref{sec:nested}.

Some additional care has to be taken when using the approximation~\eqref{approximation} for the sign function.
If $A$ has eigenvalues close to the discontinuity along the imaginary axis, the Krylov subspace would have to be taken very large to achieve a good accuracy with Eq.~\eqref{approximation}.
This can be resolved by treating these \textit{critical eigenvalues} exactly using deflation \cite{Bloch:2007aw}. 
Assume that we have computed $m$ such critical eigenvalues $\lambda_i$ with their corresponding right and left eigenvectors $r_i$ and $l_i$, then
the function evaluation is rewritten as
$ f(A)x = R_m f(\Lambda_m) (L_m^\dagger x) + f(A)(I-R_m L_m^\dagger)x$, where $\Lambda_m=\diag(\lambda_1,\ldots,\lambda_m)$, $R_m  = \left[r_1 | \dots | r_m\right]$ and $L_m  = \left[l_1 | \dots | l_m\right]$.
The first contribution to $f(A)x$ is computed exactly using the eigensolutions for the deflated eigenvalues, while the second contribution is approximated using the Krylov subspace approximation \eqref{approximation}. 

\begin{figure}[b]
\setlength{\figwidth}{0.45\textwidth}
\centering
\includegraphics[width=\figwidth]{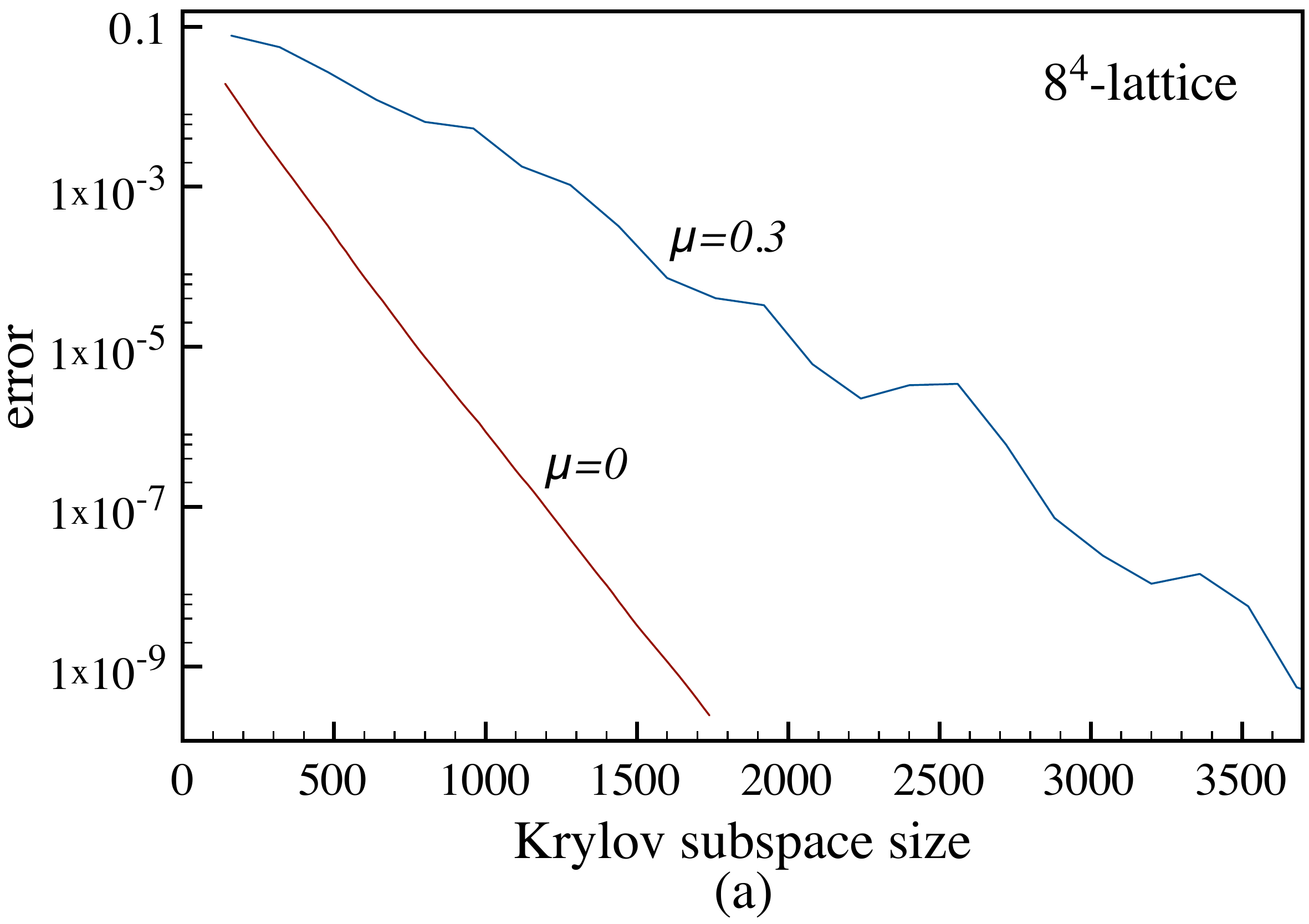}
\hspace{8mm}
\includegraphics[width=\figwidth]{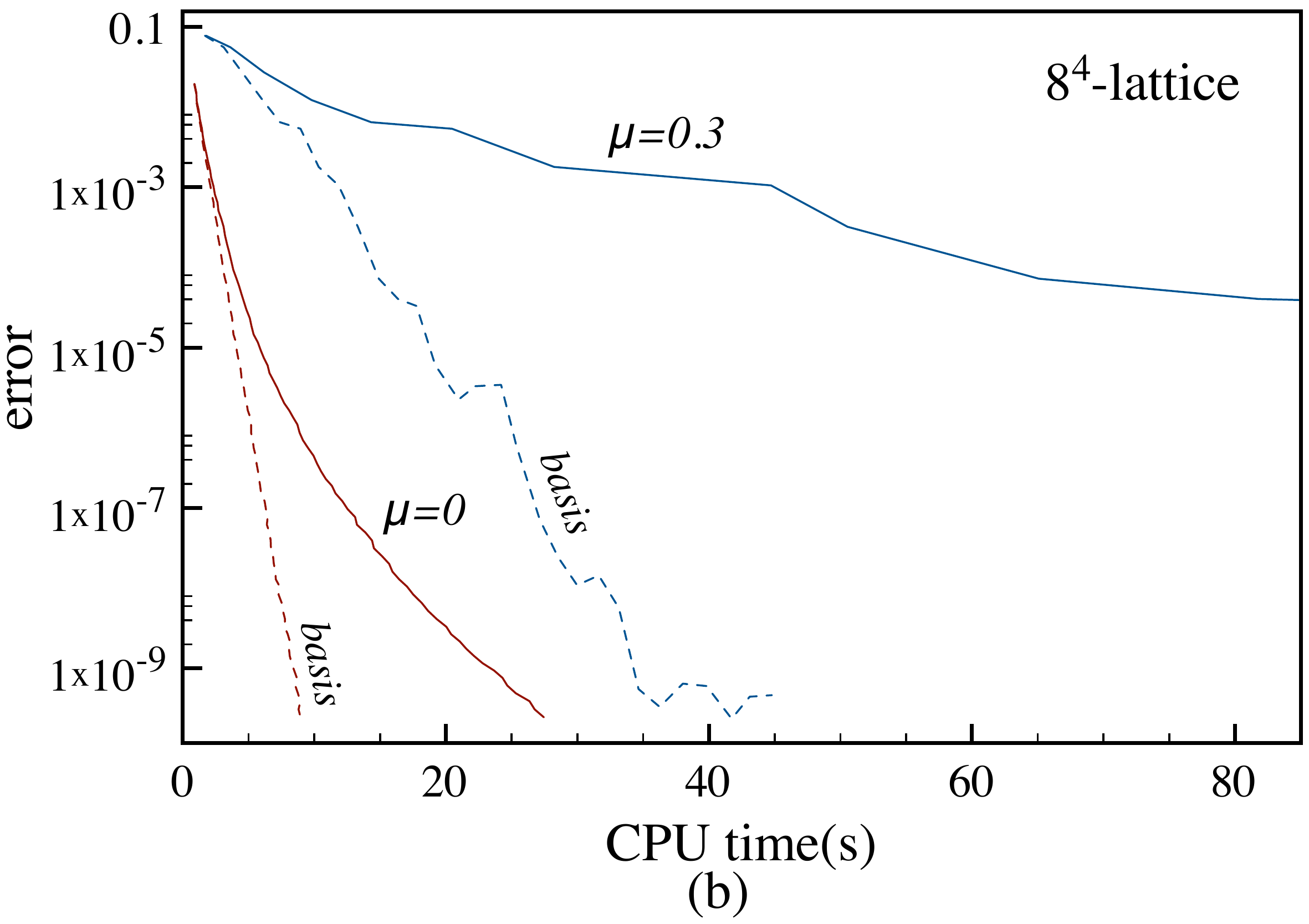}
\caption{Accuracy versus (a) Krylov subspace size (left) and (b) CPU time (right) for a typical $8^4$ configuration in the Hermitian ($\mu=0$) and non-Hermitian ($\mu=0.3$) cases, with deflation gap $\Delta = 0.055$. In the right panel the full lines show the total time required to compute $\sgn(A) x$, while the dashed lines give the time needed to build the bases in the Krylov subspaces. 
}
\label{fig:cpu}
\end{figure}
The non-nested approximation has been implemented and tested on configurations ranging from $4^4$ to $16^3 \times 32$ lattices \cite{Bloch:2007aw,Bloch:2008gh,Bloch:2009in}.
The performance of the method is illustrated for a typical $8^4$ configuration in Fig.~\ref{fig:cpu}. 
As expected, the convergence is generically faster and smoother in the Hermitian case (Lanczos) than in the non-Hermitian case (2sL), see Fig.~\ref{fig:cpu}a.
More relevant for the current study is Fig.~\ref{fig:cpu}b which depicts the CPU times used in both cases. The figure highlights a serious issue with the non-nested method, as  
a bottleneck in the numerical computation can clearly be identified, in both the Hermitian and non-Hermitian case. 
The difference between the full and dashed lines corresponds to the time needed to compute $\sgn(H_k)$ using the iteration~\protect\eqref{Roberts}.
Evidently, this computation takes up a substantial amount of the total CPU time,
as the $\mathcal{O}(k^3)$ complexity makes the Roberts-Higham iteration overly expensive for large Krylov subspaces. 
In this talk we will present a new method which alleviates this problem by computing $\sgn(H_k) e_1^{(k)}$ of Eq.~\eqref{approximation} using an additional Krylov subspace level.

\section{Nested Krylov subspace method for the sign function}
\label{sec:nested}

The idea is to approximate the vector $\sgn(H_{\kout})e_1^{(\kout)}$ of dimension $\kout$ in a smaller, nested Krylov subspace 
$\mathcal{K}_{\kin}(H_{\kout},e_1^{(\kout)})$ of dimension $\kin$ using the approximation \eqref{approximation}, with $A \to H_k$ and $x \to e_1^{(\kout)}$:
\begin{align}
  \sgn(A)x \approx \lVert x\rVert V_{\kout} \sgn(H_{\kout}) e_1^{(\kout)} 
  \approx \lVert x\rVert V_{\kout} V_{\kin} \sgn(H_{\kin}) e_1^{(\kin)} ,
\label{nested}
\end{align}
where $\sgn(H_{\kin})$ is computed using Eq.~\eqref{Roberts}.
As it stands, Eq.~\eqref{nested} does not improve the efficiency of the original method because the inner Krylov subspace $\mathcal{K}_{\kin}(H_{\kout},e_1^{(\kout)})$ only contains information coming from the $\kin \times \kin$ upper left corner of $H_{\kout}$, due to the tridiagonal nature of $H_k$ and the special source vector $e_1^{(\kout)}$. 
Therefore, the size of the inner Krylov subspace has to be chosen $\kin \approx \kout$ in order to achieve the full accuracy of the outer Krylov subspace, and nothing has been gained.

\begin{figure}[h]
\centering
\includegraphics[width=0.55\textwidth]{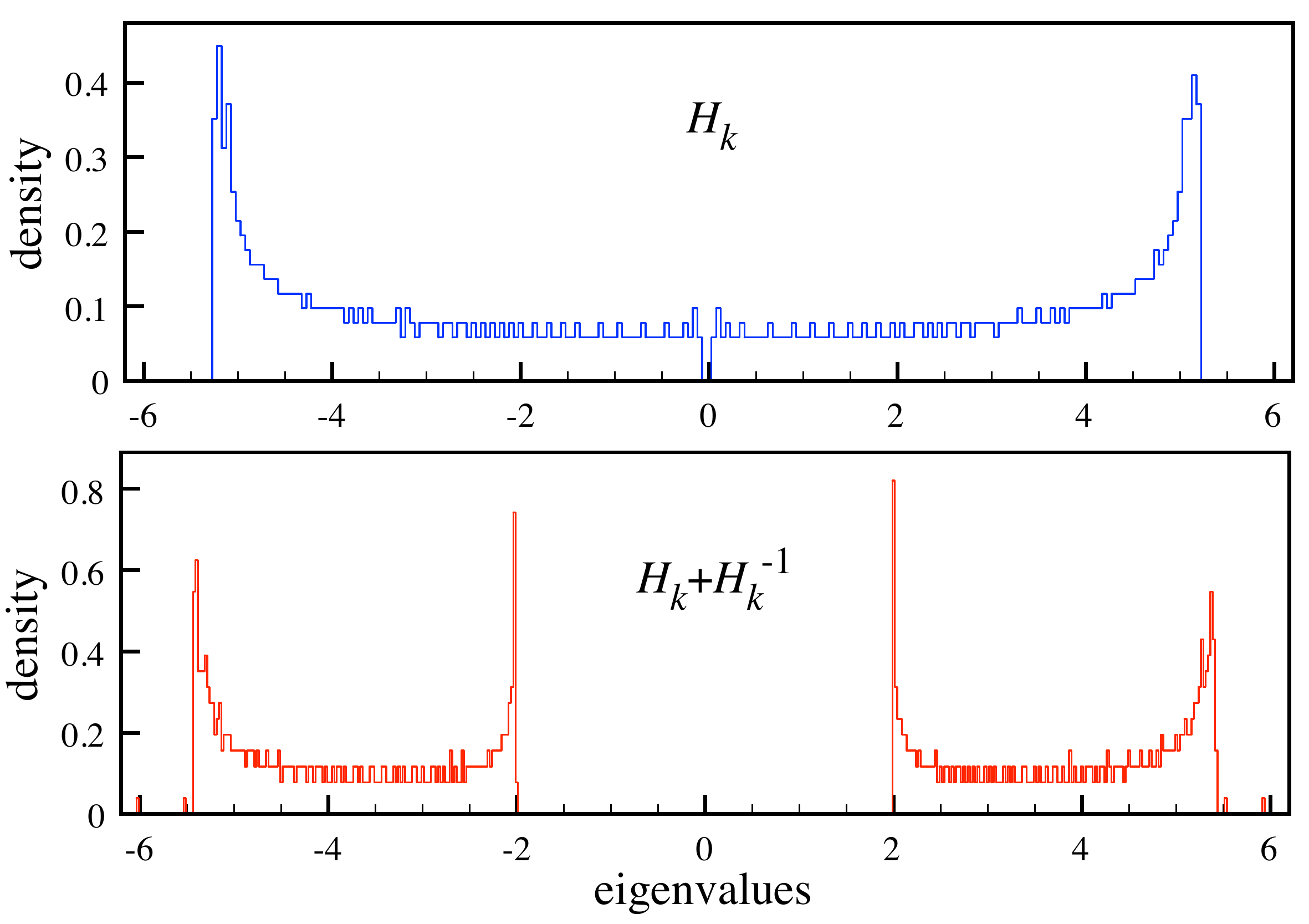}
\caption{Eigenvalue distributions of $H_{\kout}$ (top) and $H_{\kout} + H_{\kout}^{-1}$ (bottom) for a $6^4$ lattice with $k=1024$. 
}
\label{fig:spectra}
\end{figure}

There is, however, a way to circumvent this problem and make the idea work. We observe that
\begin{align} 
\sgn(A) = \sgn(A+A^{-1}) ,
\label{precond}
\end{align}
 since the eigenvectors of both arguments are the same, and
 the transformation does not change the sign of the eigenvalues, i.e.,
\begin{align}
   \sgn\left(z+\frac{1}{z}\right) = \sgn\re\left(z+\frac{1}{z}\right) &= \sgn\re\left( z+\frac{z^*}{|z|^2}\right) = \left(1+\frac{1}{|z|^2}\right)\sgn\re(z) = \sgn z \,,
\end{align}
for $z\in \mathbb{C}$.
Based on the property \eqref{precond} we introduce a preconditioning step $H_{\kout} \to H_{\kout}+H_{\kout}^{-1}$ in Eq.~\eqref{approximation}, 
which enables us to use the nested approximation \eqref{nested} with inner Krylov subspace $\mathcal{K}_{\kin}(H_{\kout}+H_{\kout}^{-1},e_1^{(\kout)})$ instead of $\mathcal{K}_{\kin}(H_{\kout},e_1^{(\kout)})$. 
As we will see in Sec.~\ref{sec:numres}, the nested approximation with preconditioning step works well with $\kin \ll \kout$, i.e., there is no noticeable loss in accuracy even after reducing the Krylov subspace size substantially.
This increased efficiency can be understood by noting that the transformation $H_{\kout} \to H_{\kout} + H_{\kout}^{-1}$ improves the condition number by a factor ten, approximately. This is illustrated in
Fig.~\ref{fig:spectra} showing the spectra of $H_{\kout}$ and $H_{\kout} + H_{\kout}^{-1}$ for the Hermitian case. Clearly, the preconditioning step considerably widens the gap around the origin, causing the improvement in condition number. 

\section{Numerical results}
\label{sec:numres}
    
\setlength{\figwidth}{0.44\textwidth}
In this section we show some preliminary numerical results for the nested Krylov subspace method.
\begin{figure}[tbp]
\centering
\includegraphics[width=\figwidth]{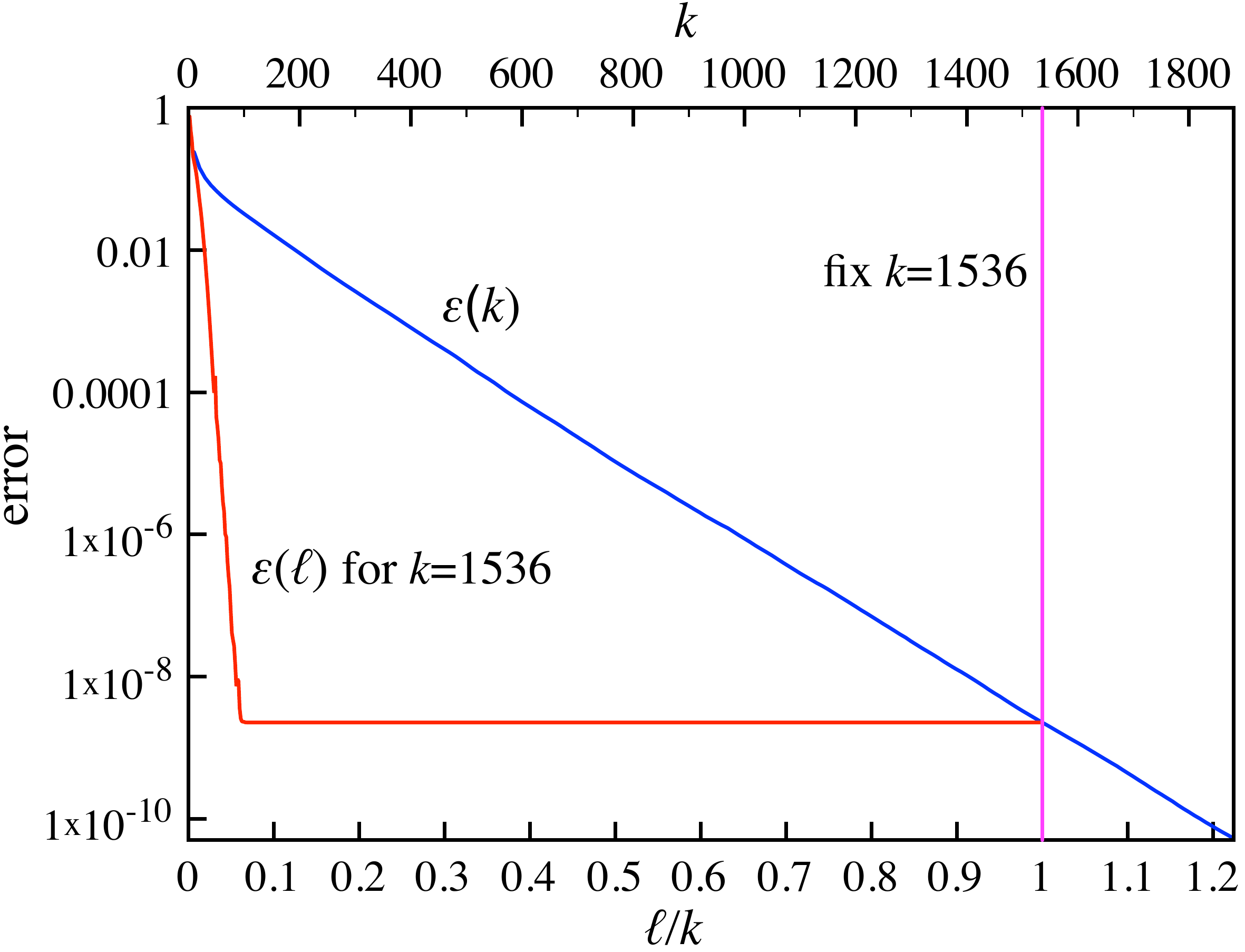}
\caption{Convergence of the nested method for an $8^4$ lattice in the Hermitian case (deflation gap $\Delta=0.055$). The blue line shows the accuracy $\varepsilon$ 
of the non-nested approximation
versus the outer Krylov subspace size $\kout$. 
The vertical line picks a value for $\kout$, here $\kout=1536$, corresponding to a desired accuracy. 
Finally, the red line shows the accuracy of the nested method as a function of the inner Krylov subspace size $\kin$ for fixed $\kout$. }
\label{fig:effic}
\end{figure}
The gain in efficiency of the method is characterized by the smallest value of $\kin/\kout$ for which no relevant loss of accuracy occurs.
This is illustrated in Fig.~\ref{fig:effic}, which shows the convergence of the nested method. 
The crucial feature is that, for fixed $\kout$, the accuracy of the nested approximation remains optimal over a very large range in $\kin$ until the error eventually blows up when the inner Krylov subspace becomes too small.
From the figure one observes that $\kin$ can be chosen about ten times smaller than $\kout$ without affecting the accuracy of the approximation.
\begin{figure}[b]
\centering
\includegraphics[width=\figwidth]{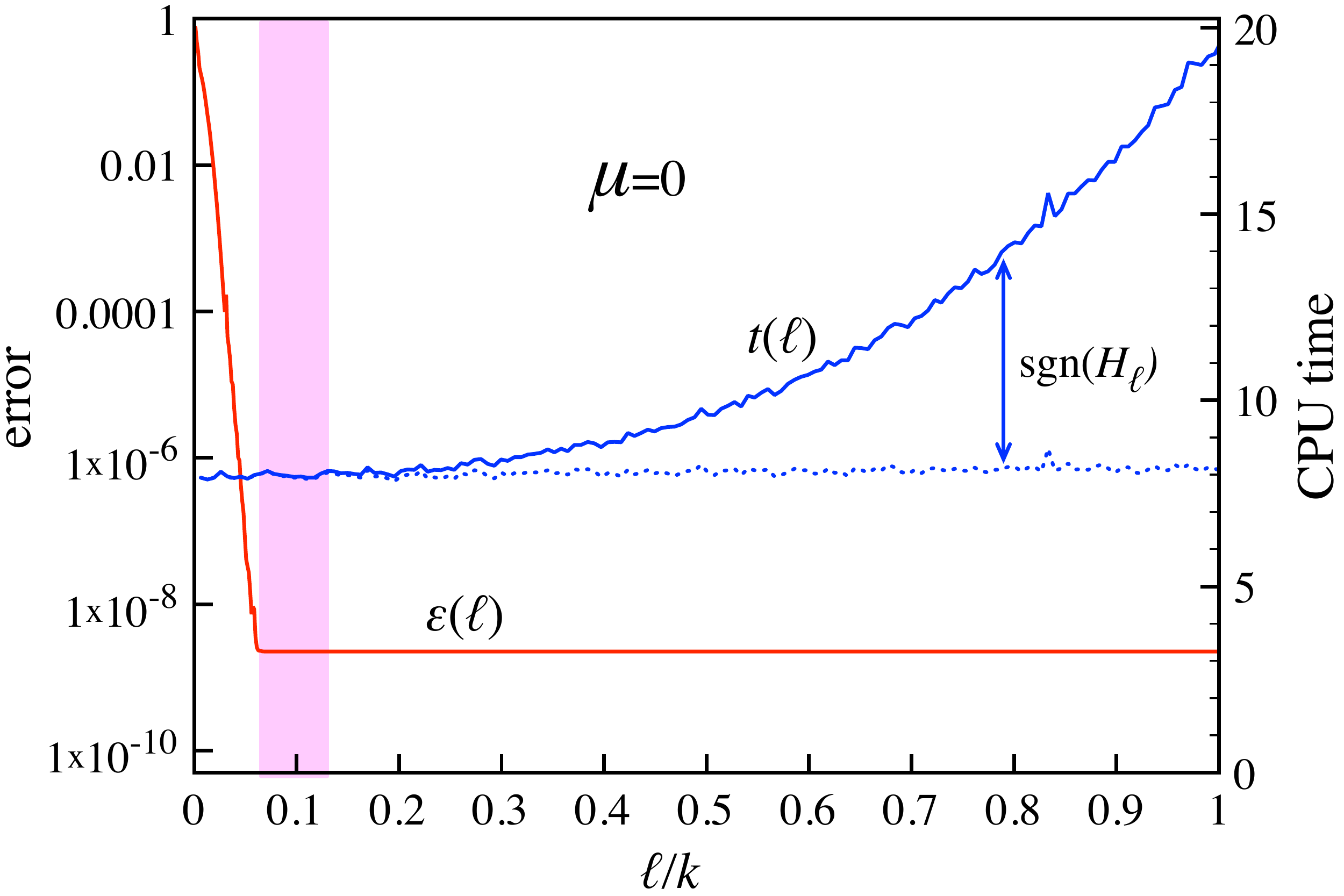}
\hspace{5mm}
\includegraphics[width=\figwidth, type=pdf,ext=.pdf,read=.pdf]{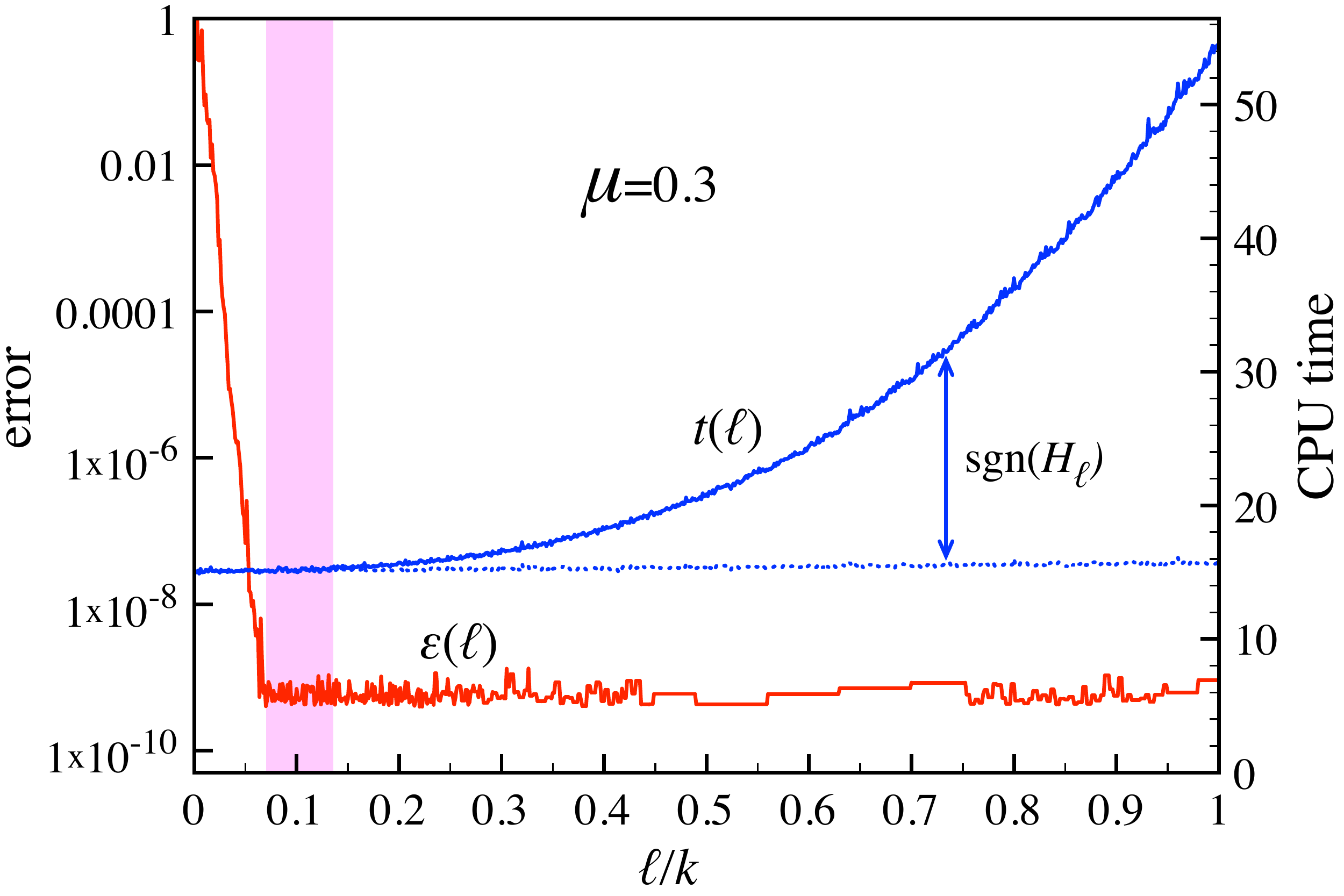}
\caption{Gain in CPU time of the nested method for an $8^4$ lattice in the Hermitian case (left) with $\kout=1536$ ($\Delta=0.055$) and the non-Hermitian case (right) for $\mu=0.3$ with $\kout=1600$ ($\Delta=0.11$). 
The red line shows the accuracy $\varepsilon(\kin)$, for a fixed  
outer Krylov subspace size $\kout$.
The total CPU time used by the method is given by the full blue line,
while the horizontal dotted line represents the time needed to construct the outer Krylov subspace. 
The difference between both corresponds to the time needed to compute $\sgn(H_{\kin})$ using the Roberts-Higham iteration and is of order ${\cal O}(\kin^3)$. 
The vertical band shows the range of optimal $\kin$.
}
\label{fig:gain}
\end{figure}
Fig.~\ref{fig:gain} illustrates how this reduction translates in a gain in CPU time.
For fixed $\kout$, the inner Krylov subspace size $\kin$ is varied and the corresponding accuracy and CPU time can be read from the figure.
The large gain in CPU time achieved when reducing $\kin$ is due to the ${\cal O}(\kin^3)$ cost to compute $\sgn(H_{\kin})$. 
It is remarkable that there is a region of $\kin$, given by the vertical band, where the accuracy is still maximal but where the computation time of $\sgn(H_{\kout})e_1^{(\kout)}$ is negligible. 
This makes the nested method extremely efficient, at least for the lattice sizes considered in these preliminary tests.

\setlength{\figwidth}{0.41\textwidth}
\begin{figure}[b]
\centering
\includegraphics[width=\figwidth,type=pdf,ext=.pdf,read=.pdf]{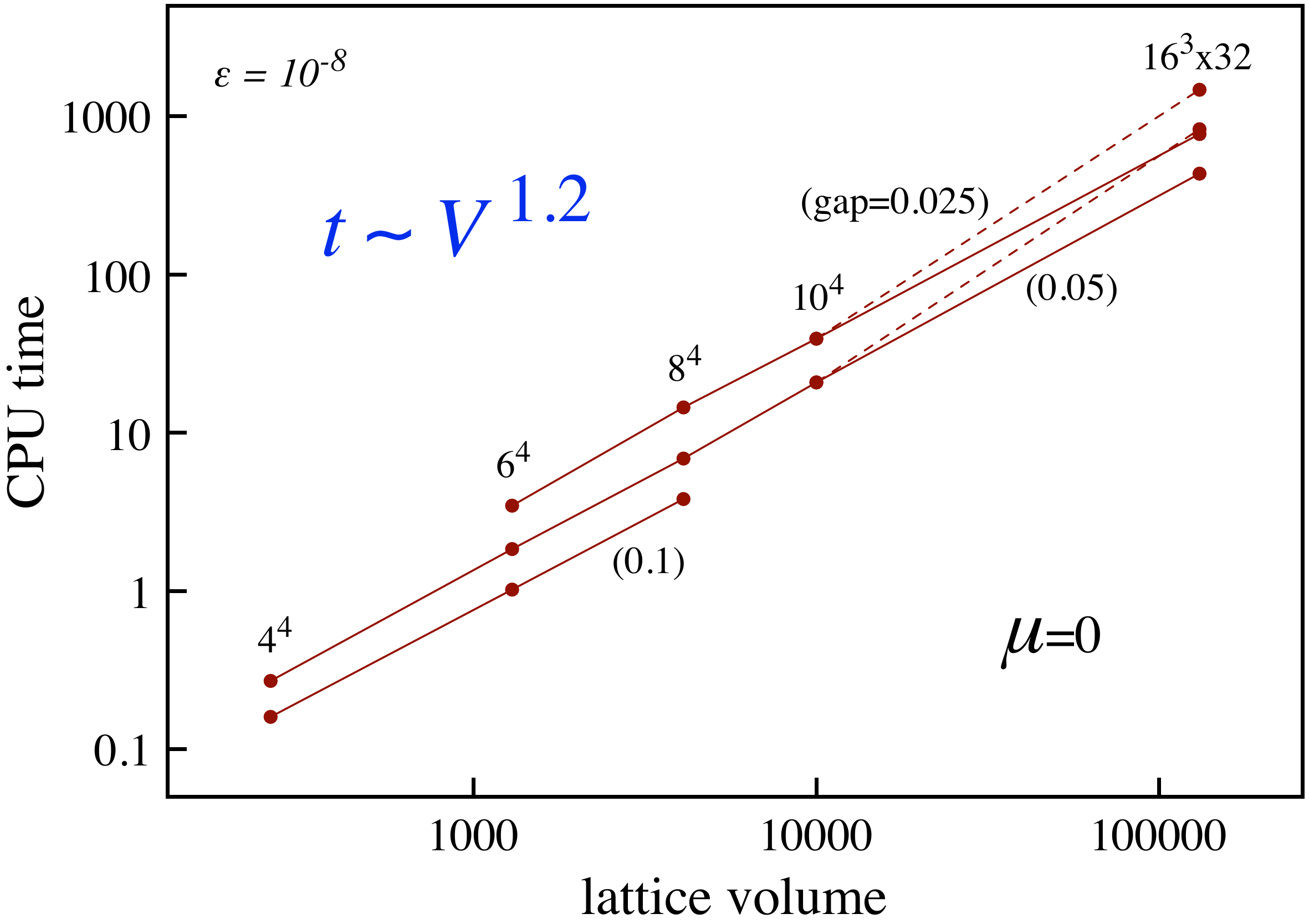}
\hspace{10mm}
\includegraphics[width=\figwidth,type=pdf,ext=.pdf,read=.pdf]{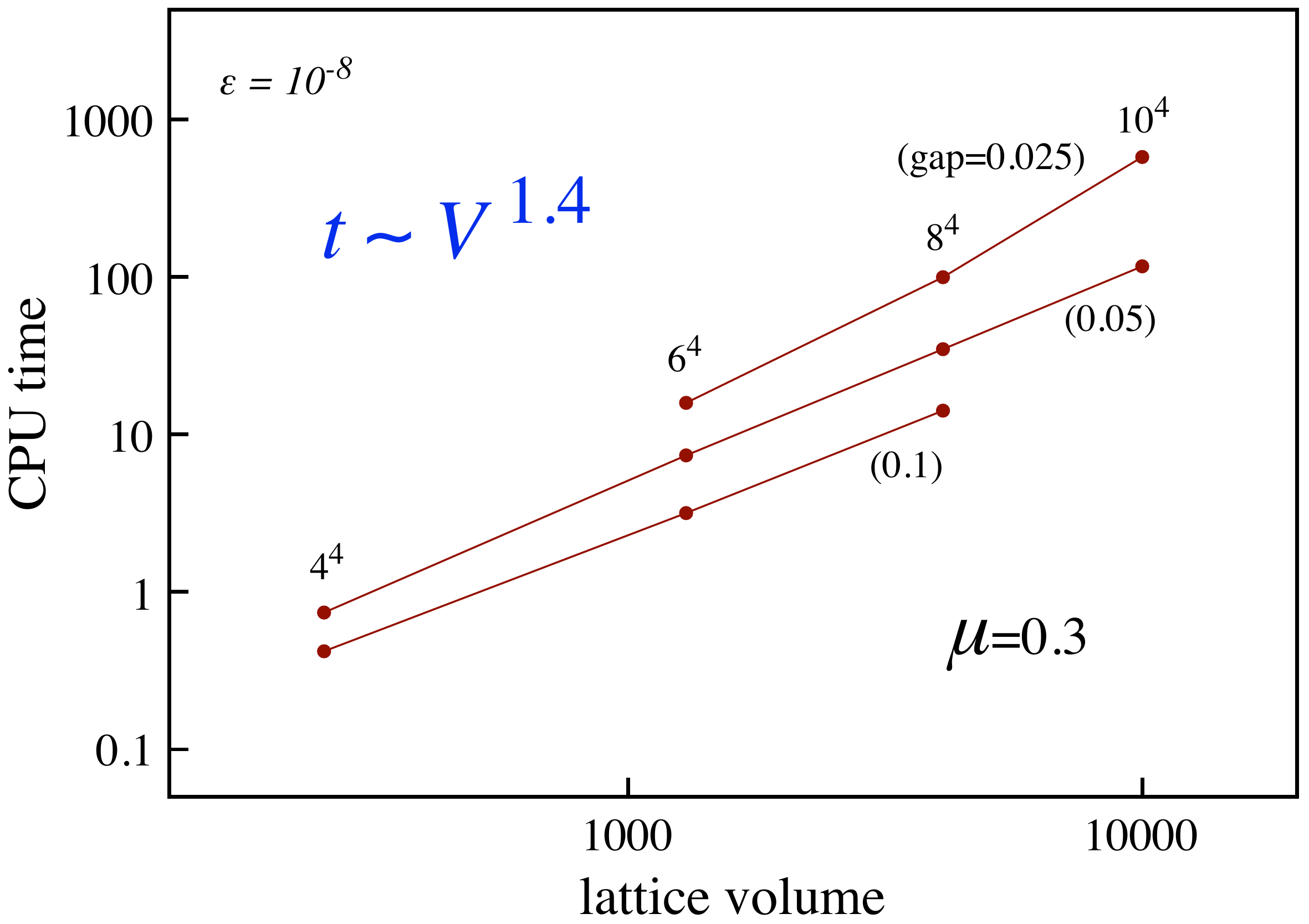}
\caption{Scaling with matrix size: CPU time versus lattice volume for the Hermitian case (left) and non-Hermitian case with $\mu=0.3$ (right) for three different deflation gaps: 0.025, 0.5 and 0.1. The requested accuracy is $\epsilon=10^{-8}$. The dashed lines represent double pass calculations \cite{S&J:2009}.}
\label{fig:scaling}
\end{figure}
To explore the employability of the method for realistic lattice calculations it is useful to investigate how the algorithm scales with the lattice volume. This is illustrated in Fig.~\ref{fig:scaling}. Fitting the CPU time gives an approximate volume dependence $t \sim V^{1.2}$ for the Hermitian case and $t \sim V^{1.4}$ for the non-Hermitian case.
\begin{figure}[b]
\centering
\includegraphics[width=\figwidth]{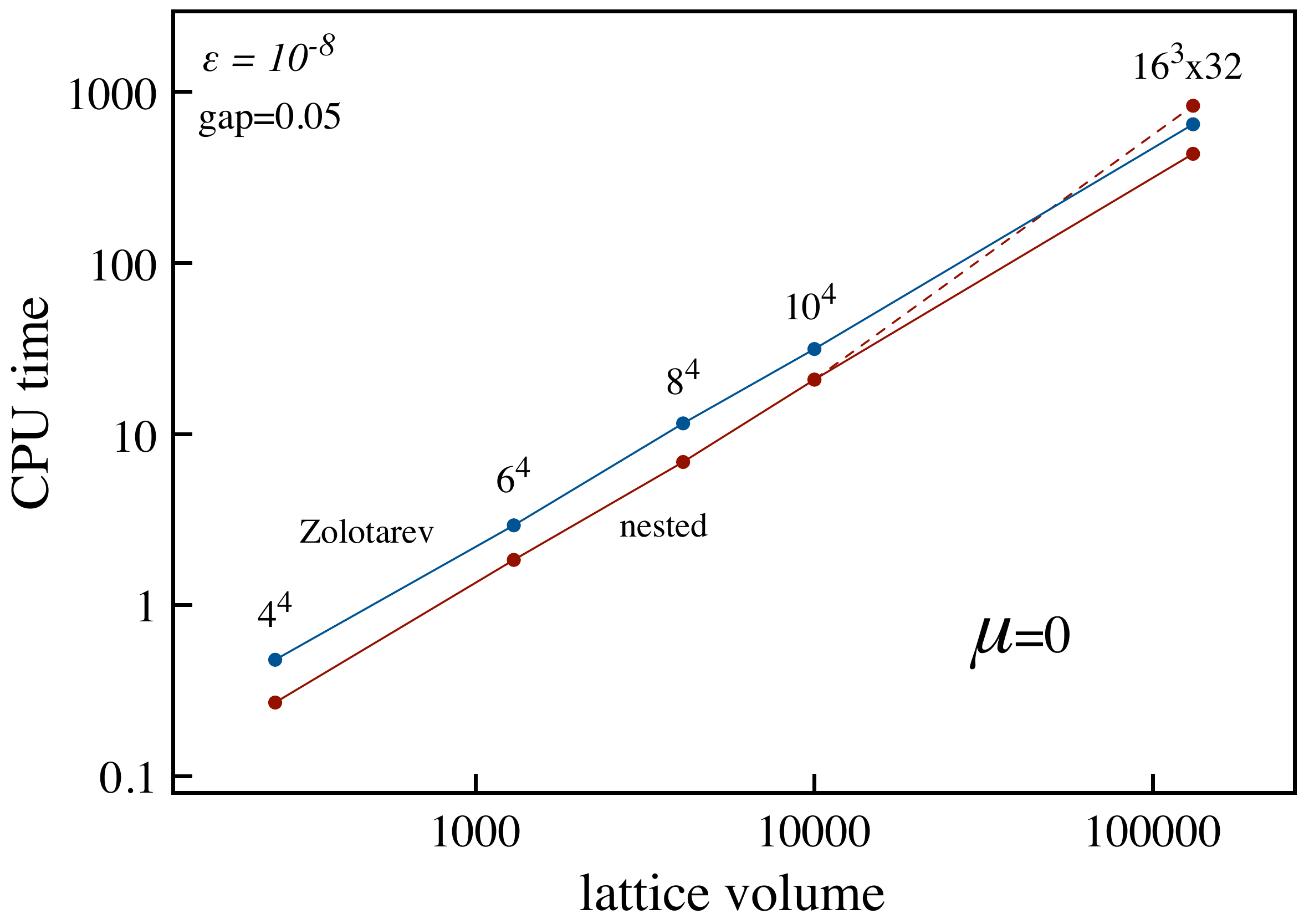} 
\hspace{10mm}
\includegraphics[width=\figwidth]{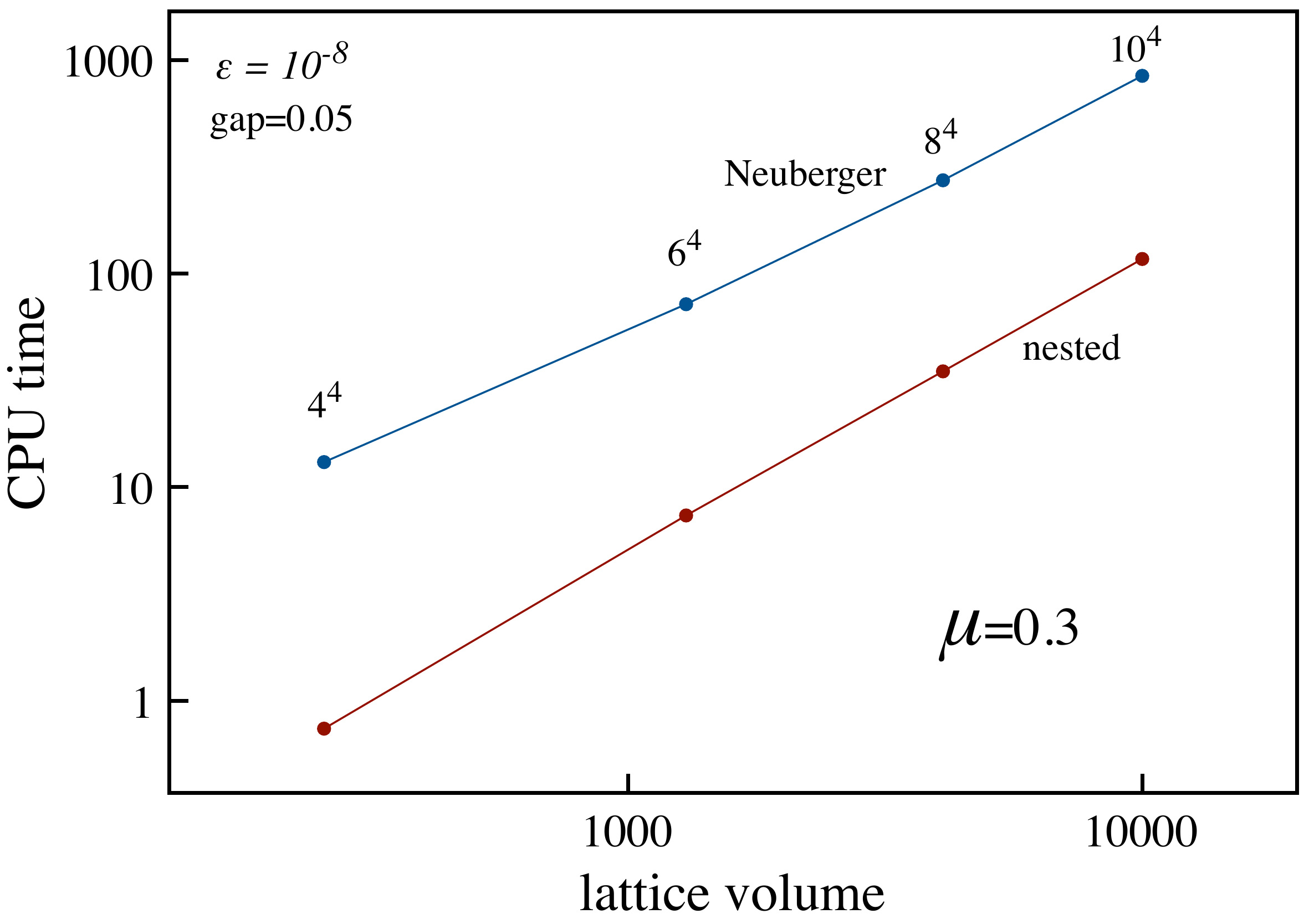} 
\caption{Comparison of the CPU time used by the nested method and rational approximation methods: For the Hermitian case (left) we compare the nested method with the Zolotarev approximation \cite{vandenEshof:2002ms},
for the non-Hermitian case (right) we compare with the Neuberger approximation computed with a multishift, restarted FOM method \cite{Bloch:2009in}. The requested accuracy is  $\epsilon=10^{-8}$ and the deflation gap is $\Delta=0.05$.}
\label{fig:compare}
\end{figure}
In this context we also compare the efficiency of our new method with the rational approximation methods, which are the best methods currently on the market, see Fig.~\ref{fig:compare}.
For the Hermitian case the nested method performs slightly better than the Zolotarev method. 
However, the difference is not significant and conclusions may depend on the details of the implementation 
(especially since the nested method is sometimes used in double pass mode, depending on the available hardware, to avoid storage problems for large lattices \cite{S&J:2009}). 
For the non-Hermitian case, where we compare with the rational approximation method presented in Ref.~\cite{Bloch:2009in}, the conclusion is unambiguous and the nested method is by far better than the rational approximation. 
This is caused by the lesser efficiency of the rational approximation in the presence of complex eigenvalues \cite{Bloch:2009in}.

\section{Summary and outlook}

The Ritz approximation to the sign function slows down dramatically when the Krylov subspace grows large.
We therefore developed an improvement based on nested Krylov subspaces, which resolves this problem and expedites the computation of the sign function for Hermitian and non-Hermitian matrices, without affecting the accuracy of the approximation.
Moreover, the new method turns out to be a worthy alternative to state-of-the-art rational approximation methods.
More details about the nested method can be found in Ref.~\cite{S&J:2009}.

Future developments will include the parallel implementation and benchmarking of the nested method, its incorporation in hybrid Monte Carlo algorithms for dynamical simulations with overlap fermions, and the investigation of the applicability of the nested method to other matrix functions.

\acknowledgments

We would like to thank Andreas Frommer and Tilo Wettig for discussions.

\bibliographystyle{JHEP}
\bibliography{/Users/bloch/physics/notes/biblio}

\providecommand{\href}[2]{#2}\begingroup\raggedright\begin{thebibliography}{10}

\bibitem{S&J:2009}
S.~Heybrock and J.~C.~R. Bloch, {\it A nested {K}rylov subspace method to
  compute the sign function of a complex matrix},  in preparation.

\bibitem{Narayanan:1994gw}
R.~Narayanan and H.~Neuberger, {\it {A construction of lattice chiral gauge
  theories}},  {\em Nucl. Phys.} {\bf B443} (1995) 305--385,
  [\href{http://xxx.lanl.gov/abs/hep-th/9411108}{{\tt hep-th/9411108}}].

\bibitem{Neuberger:1997fp}
H.~Neuberger, {\it {Exactly massless quarks on the lattice}},  {\em Phys.
  Lett.} {\bf B417} (1998) 141--144,
  [\href{http://xxx.lanl.gov/abs/hep-lat/9707022}{{\tt hep-lat/9707022}}].

\bibitem{Bloch:2006cd}
J.~C.~R. Bloch and T.~Wettig, {\it {Overlap Dirac operator at nonzero chemical
  potential and random matrix theory}},  {\em Phys. Rev. Lett.} {\bf 97} (2006)
  012003, [\href{http://xxx.lanl.gov/abs/hep-lat/0604020}{{\tt
  hep-lat/0604020}}].

\bibitem{Hasenfratz:1983ba}
P.~Hasenfratz and F.~Karsch, {\it Chemical potential on the lattice},  {\em
  Phys. Lett.} {\bf B125} (1983) 308.

\bibitem{Bloch:2007xi}
J.~C.~R. Bloch and T.~Wettig, {\it {Domain-wall and overlap fermions at nonzero
  quark chemical potential}},  {\em Phys. Rev.} {\bf D76} (2007) 114511,
  [\href{http://xxx.lanl.gov/abs/0709.4630}{{\tt arXiv:0709.4630}}].

\bibitem{Rob80}
J.~Roberts, {\it Linear model reduction and solution of the algebraic {Riccati}
  equation by use of the sign functions},  {\em Internat. J. Control} {\bf 32}
  (1980) 677--687.

\bibitem{Bloch:2007aw}
J.~C.~R. Bloch, A.~Frommer, B.~Lang, and T.~Wettig, {\it {An iterative method
  to compute the sign function of a non- Hermitian matrix and its application
  to the overlap Dirac operator at nonzero chemical potential}},  {\em Comput.
  Phys. Commun.} {\bf 177} (2007) 933--943,
  [\href{http://xxx.lanl.gov/abs/0704.3486}{{\tt arXiv:0704.3486}}].

\bibitem{Bloch:2008gh}
J.~C.~R. Bloch, T.~Breu, and T.~Wettig, {\it {Comparing iterative methods to
  compute the overlap Dirac operator at nonzero chemical potential}},  {\em
  PoS} {\bf LATTICE2008} (2008) 027,
  [\href{http://xxx.lanl.gov/abs/0810.4228}{{\tt arXiv:0810.4228}}].

\bibitem{Bloch:2009in}
J.~C.~R. Bloch, T.~Breu, A.~Frommer, S.~Heybrock, K.~Sch\"afer, and T.~Wettig,
  {\it {Short-recurrence Krylov subspace methods for the overlap Dirac operator
  at nonzero chemical potential}},
  \href{http://xxx.lanl.gov/abs/0910.1048}{{\tt arXiv:0910.1048}}.

\bibitem{vandenEshof:2002ms}
J.~van~den Eshof, A.~Frommer, T.~Lippert, K.~Schilling, and H.~A. van~der
  Vorst, {\it {Numerical methods for the QCD overlap operator. I: Sign-
  function and error bounds}},  {\em Comput. Phys. Commun.} {\bf 146} (2002)
  203--224, [\href{http://xxx.lanl.gov/abs/hep-lat/0202025}{{\tt
  hep-lat/0202025}}].

\end{thebibliography}\endgroup

\end{document}